\documentstyle{article}
\title{Universal regular short distance behavior from
an interaction with a scale invariant gravity}
\author{ Z. Haba\\Institute of Theoretical Physics, University of Wroclaw,
Poland\\e-mail:zhab@ift.uni.wroc.pl}
\date{}
\begin{document}
\maketitle
\begin{abstract}
We assume that the fourdimensional quantum gravity is scale invariant
at short distances. We show through
a simple scaling argument that
correlation functions of quantum fields interacting
with gravity  have a universal (more regular)
short distance behavior.

\end{abstract}
We consider the tetrads $e^{\mu}_{a}(x)$ as the basic variables
 for quantum gravity.
We assume that
correlation functions of
$e^{\mu}_{a}(x)$ are scale invariant.
This means that (with a certain real $\gamma$) $\lambda^{\gamma}e^{\mu}_{a}(\lambda x )$
and $e^{\mu}_{a}(x)$ have the same correlation functions
for all real $\lambda$,i.e.,
these are equivalent random variables.
Moreover, we assume that at short distances
$ <e_{a}^{\mu }( x)e_{c}^{\nu}( x^{\prime})>$
depends only on $x-x^{\prime}$. The scale invariance
cannot apply to
 intermediate distances where the
correspondence with the classical gravity is required.
Then, we  expect that the mean value of $e^{\mu}_{a}(x)$
describes the classical gravity. We could apply our arguments
if  the mean value of the tetrad was different
from zero and the correlation functions were scale invariant only
for short distances. However, for simplicity of the
presentation we assume the exact scale invariance.
We show that  quantum fields interacting with
gravity have a more regular behavior than the canonical one.
It is
determined solely by the scaling index $\gamma$.
In reality  quantum gravity is not expected
to be exactly scale invariant. There will be a distance scale
(presumably the Planck scale) which determines the
region where the scaling applies.
 There are renormalizable models of gravity (with the square of the curvature tensor)
 \cite{stelle}\cite{julve} \cite{tomb}\cite{frad} which may be scale
 invariant at short distances.
 Experiments support
the canonical behavior of QED
at short distances (modified eventually by logarithmic
corrections). We suggest that the regular (superrenormalizable)
behavior takes over at distances below the Planck scale
( there were earlier papers on this topic reviewed by S. Deser
in \cite{deser}, see also \cite{salam}\cite{ford}).

We consider first the free scalar field
in four dimensions interacting
with gravity. The two-point function in an external
gravitational field is
(in the proper time representation)
\begin{equation}
\begin{array}{l}
G_{g}(x,y)\equiv\int_{0}^{\infty}d\tau K_{\tau}(x,y)=
\int_{0}^{\infty}d\tau\int {\cal D}x\exp(-\frac{1}{2}\int g^{\mu\nu}\frac{dx_{\mu}}{dt}
  \frac{dx_{\nu}}{dt})
 \cr
 \delta\left(x\left(0\right)-x\right)
  \delta\left(x\left(\tau\right)-y\right)
  \end{array}
  \end{equation}
We can transform the functional integral (1) into a Gaussian
integral
by means of a change of variables
$x(s)\rightarrow b(s)$. For this purpose we
choose $e^{\mu}_{a}(x)$ as a square root of the metric
(defined up to a rotation)
\begin{equation}
g^{\mu\nu}(x)=e^{\mu}_{a}(x)e^{\nu}_{a}(x)
\end{equation}
Then, we define a path  $q^{\mu}(s)$ starting from $x$
and the tetrad $e$ along this path by the system of
(Stratonovitch) equations
(see \cite{ikeda} or \cite{halpern}\cite{haba} for a heuristic derivation)
\begin{equation}
dq^{\mu}(s)=e_{a}^{\mu}\left( q\left(s\right)\right)db^{a}(s)
\end{equation}
\begin{equation}
de^{\mu}_{a}(q)+\Gamma^{\mu}_{\nu\rho}(q)e^{\nu}_{a}(q)dq^{\rho}=0
\end{equation}
where $\Gamma$ is the Christoffel symbol.
It follows from eq.(1) that $b^{a}(s)$ is
a Gaussian process (the Brownian motion) with the
covariance (the expectation value over the Brownian motion will
be denoted by $E[.]$)
\begin{displaymath}
E[b_{a}(t)b_{c}(s)]=\delta_{ac}\min(s,t)
\end{displaymath}
When we solve eqs.(3)-(4) then the heat kernel
entering eq.(1) is expressed by the formula
\begin{equation}
\begin{array}{l}
K_{\tau}(x,y)=E[\delta(y-q_{\tau}(x))]
\end{array}
\end{equation}
The short distance behavior of the two-point function
(1) is determined by the behavior of $q(\tau)-x$ for a small $\tau$.
We can obtain this behavior from the differential equation (3).
Let us assume (here the equivalence is meant in the sense of
correlation functions)
\begin{equation}
q(\alpha s)-x\simeq \alpha^{\omega}(\tilde{q}(s)-x)
\end{equation}
where $\tilde{q}$ is equivalent to $q$ .
Then, using
\begin{equation}
\lambda^{\gamma}e(\lambda x)\simeq \tilde{e}(x)
\end{equation}
(where $\tilde{e}(x)$ is a random field
equivalent to $e(x)$) and the (approximate) translational invariance of the correlation
 functions of the tetrads  we can determine $\omega$
 inserting eqs.(6)-(7) into eq.(3)
 \begin{equation}
 \omega=\frac{1}{2}(1+\gamma)^{-1}
 \end{equation}
 Eq.(7) means that $e(x)e(y)\approx \vert x-y\vert^{-2\gamma}$
 (in the sense of correlation functions). Hence, the singular short
 distance behavior of the random field $e$ makes
 the paths $q$ less regular (we would have $\omega=\frac{1}{2}$
 for the Brownian motion as well as  for  regular tetrads).

 In quantum gravity the scalar propagator $G$ results
 from an average $\langle G_{g}\rangle $ of $G_{g}$ in eq.(1) over the
 gravitational field. In order to calculate this average
 let us write eq.(3) in an integral form
 \begin{equation}
 q^{\mu}(\tau)=x^{\mu}+\int_{0}^{\tau}e^{\mu}_{a}
 \left(q\left(s\right)\right)db^{a}(s)
 \end{equation}
Inserting the r.h.s. of eq.(9) into  eq.(5) we obtain
\begin{equation}
\begin{array}{l}
\langle K_{\tau}(x,y)\rangle=\langle E\left[\delta\left(y-x-\int_{0}^{\tau}
e^{\mu}_{a}
 \left(q\left(s\right)\right)db^{a}\left(s\right)
\right)\right]\rangle
\end{array}
\end{equation}
We consider a change of time $s\rightarrow \frac{s}{\tau}$
inside the integral (10) and  apply the scaling (7) of $e$,
the scaling $b(\alpha s)\simeq \sqrt{\alpha}b(s)$ of the Brownian
motion and the scaling
(6) of $q(s)$  (with $\alpha=\frac{1}{\tau}$)
 in order to rewrite the kernel (10) in terms of the paths of the
 process $\tilde{q}$ defined on the interval $[0,1]$
\begin{equation}
\begin{array}{l}
\langle K_{\tau}(x,y)\rangle=\langle E\left[\delta\left(y-x-
\tau^{\frac{1}{2}-\omega\gamma}\int_{0}^{1}
\tilde{e}^{\mu}_{a}
 \left(\tilde{q}\left(s\right)\right)db^{a}\left(s\right)
\right)\right]\rangle
\cr
=\tau^{-2+4\omega\gamma}
\langle E\left[\delta\left(\left(y-x\right)\tau^{-\frac{1}{2}+\omega\gamma}
-\int_{0}^{1}
\tilde{e}^{\mu}_{a}
 \left(\tilde{q}\left(s\right)\right)db^{a}\left(s\right)
\right)\right] \rangle
\end{array}
\end{equation}
Let $F(u)$ be the probability distribution of
the random variable
\begin{displaymath}
\eta^{\mu}=\int_{0}^{1}
\tilde{e}^{\mu}_{a}
 \left(\tilde{q}\left(s\right)\right)db^{a}\left(s\right)
\end{displaymath}
Then, using eqs.(8) and (11)
\begin{equation}
\begin{array}{l}
G(x,y)=\langle G_{g}(x,y)\rangle=\int_{0}^{\infty}
d\tau\tau^{-2+4\gamma\omega}F\left(\left( y-x\right)
\tau^{-\frac{1}{2}+\omega\gamma}\right)
\cr
=2(1+\gamma)\vert y-x\vert^{-2 +2\gamma}\int_{0}^{\infty}dt
t^{1-2\gamma} F\left(t
\left (y-x\right)\vert y-x\vert^{-1}\right)
\end{array}
\end{equation}
Strictly speaking correlation functions of the random variable
$\eta$ need renormalization because $e$ has singular correlation
functions. This can easily be seen already  for the second moment
of $\eta$.
Then, a subtraction  of infinities defines a renormalized
two-point function of the scalar field (in our earlier paper
\cite{brze} we have discussed such a renormalization in a
particular model with a Gaussian tetrad). However, the result (12)
does not depend on the way we renormalize $\eta$. It follows just
from the scaling of $\tau$. The behavior of $G$ is more regular
if the gravitational correlations are more singular (in
\cite{brze} we obtained a bound  $\gamma<\frac{1}{2}$ resulting
from the requirement of renormalizability).

We could treat the perturbative expansion of $\phi^{4}$ by means
of the functional representation (1). In such a case we can show
just by the scaling argument that the correlations of the fourth
power of $\phi$ have the singularity of the fourth power of the
two point function (12). The model becomes superrenormalizable.
We can repeat the argument for other fields entering the Standard
Model. We have the path integral representation of the two-point
function in the gravitational field analogous to eq.(1) (we
consider the square of the Dirac operator is such a
representation). There are factors multiplying the
$\delta$-function inside    the path integral (5). However, such
factors have no effect on the scaling argument. Hence, the
estimate (12) remains true for the propagator of the gauge fields
as well as for the propagator of the square of the Dirac operator.

\end{document}